\pdfoutput=1
\documentclass[11pt]{article}

\usepackage[]{acl}

\usepackage{times}
\usepackage{latexsym}

\usepackage[T1]{fontenc}

\usepackage[utf8]{inputenc}

\usepackage{microtype}

\usepackage{booktabs}
\usepackage{xspace}
\usepackage{graphicx}
\usepackage{subcaption}
\usepackage{enumitem}
\usepackage[normalem]{ulem}
\useunder{\uline}{\ul}{}
\usepackage{spverbatim}
\newcommand{\proposeddataset}{HumanExtension\xspace}
\newcommand{\humaneval}{HumanEval\xspace}
  
\title{Exploring Language Model's Code Generation Ability with Auxiliary Functions}

\author{Seonghyeon Lee\textsuperscript{\textdagger}, Sanghwan Jang\textsuperscript{\textdagger}, Seongbo Jang\textsuperscript{\textdagger}, Dongha Lee\textsuperscript{\textdaggerdbl}, Hwanjo Yu\textsuperscript{\textdagger}\thanks{\ \ Corresponding author}\\
Department of Computer Science and Engineering, POSTECH, Pohang, South Korea\textsuperscript{\textdagger} \\
Department of Aritificial Intelligence, Yonsei University, Seoul, South Korea\textsuperscript{\textdaggerdbl} \\
\texttt{\{sh0416,s.jang,jang.sb,hwanjoyu\}@postech.ac.kr} \\
\texttt{donalee@yonsei.ac.kr}}

\begin{document}
\maketitle
\begin{abstract}
Auxiliary function is a helpful component to improve language model's code generation ability.
However, a systematic exploration of how they affect has yet to be done.
In this work, we comprehensively evaluate the ability to utilize auxiliary functions encoded in recent code-pretrained language models.
First, we construct a human-crafted evaluation set, called \proposeddataset, which contains examples of two functions where one function assists the other.
With \proposeddataset, we design several experiments to examine their ability in a multifaceted way.
Our evaluation processes enable a comprehensive understanding of including auxiliary functions in the prompt in terms of effectiveness and robustness.
An additional implementation style analysis captures the models' various implementation patterns when they access the auxiliary function.
Through this analysis, we discover the models' promising ability to utilize auxiliary functions including their self-improving behavior by implementing the two functions step-by-step.
However, our analysis also reveals the model's underutilized behavior to call the auxiliary function, suggesting the future direction to enhance their implementation by eliciting the auxiliary function call ability encoded in the models.
We release our code\footnote{\url{https://github.com/sh0416/humanextension}} and dataset\footnote{\url{https://huggingface.co/datasets/sh0416/humanextension}} to facilitate this research direction.
\end{abstract}

\section{Introduction}
\label{sec:introduction}
Program synthesis, i.e., writing function code by taking natural language descriptions as inputs, has garnered attention in the research community \citep{yin-neubig-2017-syntactic,10.1007/978-3-030-32520-6_6,austin2021program,doi:10.1126/science.abq1158}.
With the help of language modeling, several code-pretrained Large Language Models (LLMs) implement functions with prompts that contain target function signatures \citep{fried2023incoder,nijkamp2023codegen,nijkamp2023codegen2,allal2023santacoder,li2023starcoder,gunasekar2023textbooks}. 
Additional code components, e.g., comment lines \citep{gao2023pal}, documents \citep{zhou2023docprompting}, and other function and class definitions across files \citep{ding2023cocomic}, have been attached to the prompts to boost up their implementation ability.

\begin{figure}
\centering
\includegraphics[width=\columnwidth]{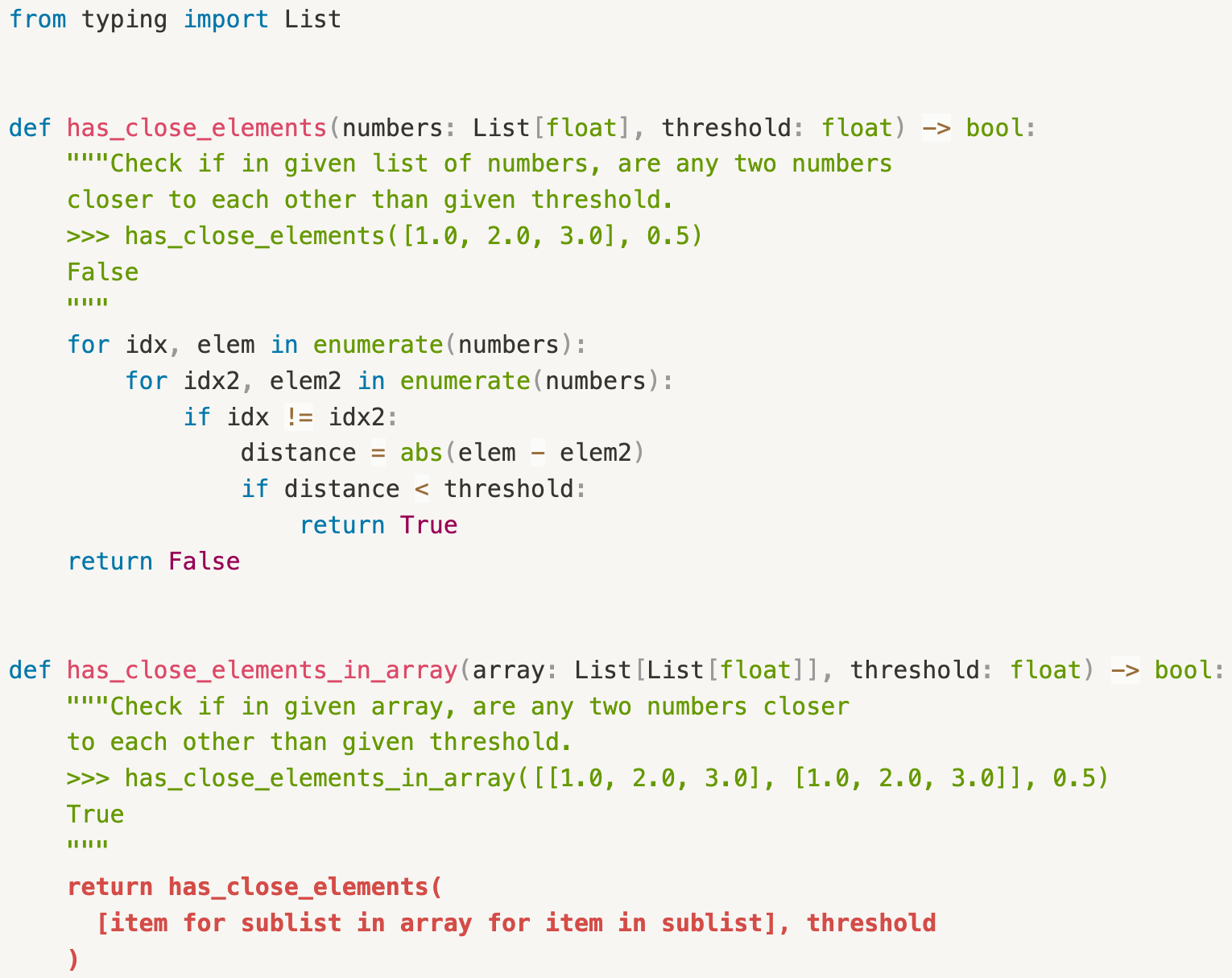}
\caption{An illustrative example of \proposeddataset. The function has\_close\_elements\_in\_array delegates their subroutine to the auxiliary function has\_close\_elements. Red bold text is the reference implementation written by humans.}
\label{fig:example_intro}
\end{figure}

Auxiliary function is one promising component to improve their code synthesis ability.
We define the auxiliary function as a function that handles a subroutine for the target one or performs an easier version of the actual requirements.
When this function is included in the prompt, LLMs could call the function to delegate their subroutine or refer to their implementation while synthesizing the target function.
However, due to the lack of an evaluation dataset that enables a systematic examination of how these auxiliary functions are utilized, no structured analysis has yet to be conducted.

In this work, we investigate several LLMs' ability to utilize auxiliary functions.
To do this, we first construct an evaluation dataset, called \proposeddataset, which contains human-crafted examples of two functions that are closely related to each other (Figure~\ref{fig:example_intro}).
Specifically, we guided labelers to extend functions in the \humaneval dataset \citep{chen2021evaluating}.
We offer software design concepts related to function extension such as subtyping \citep{liskov1994behavioral} to promote labelers to create realistic function relationships.
Additionally, the curated examples are parsed into several components to enable robustness evaluation similar to~\citet{wang-etal-2023-recode}.

With the \proposeddataset dataset, we conduct systematic analyses to understand how LLMs leverage auxiliary functions.
First, we investigate if appending a single auxiliary function to the prompt enhances the likelihood of accurately implementing the target function.
Specifically, we design several prompts with auxiliary functions while considering their existence, their functional relevance, and the availability to access auxiliary function implementations.
With these prompts, we generate implementations with LLMs and analyze the model behavior focusing on the auxiliary function's effectiveness, robustness, and the models' implementation style.
Second, we examine the cases where LLMs can access multiple auxiliary functions for synthesizing target functions.
The randomly sampled auxiliary functions are additionally included in the prompts to verify whether LLMs can selectively use the appropriate one.
Similar to \citet{liu2023lost}, we inspect whether the position of a relevant function affects their code generation ability.
This investigation is combined with the implementation style analysis to permit an in-depth analysis through the lens of the auxiliary function call.

Our experimental results show current LLMs' capabilities to utilize auxiliary function and their limitations.
First, most LLMs exhibit large performance improvement with proper relevant auxiliary functions.
Also, for some advanced LLMs, our evaluation process sheds light on their self-improving behavior by implementing the two functions in a step-by-step manner.
However, the ability to utilize auxiliary functions is varied depending on the factors that do not change their functionality, which raises a question about their robustness.
In addition, our implementation style analysis results reveal that the models prefer repeating the internal logic in the auxiliary function even when the logic can be easily handled by simply calling them.
Finally, our human preference evaluation of their style shows this disparity between model-generated implementation and that of humans, suggesting the future direction of enhancing the ability to delegate their subroutine to the auxiliary functions by calling them.

% 인용 못했는데 상관없을 듯. zelikman2023parsel,shrivastava2023repositorylevel,chen2023codet,chen2023teaching

\section{Related work}
\label{sec:related}
Several studies have been conducted to evaluate code generation ability \citep{10.1145/3520312.3534862}.
\citet{neelakantan2015neural,iyer-etal-2018-mapping} first introduce neural networks into code completion tasks and evaluate them on traditional metrics, e.g., BLEU. 
\citet{chen2021evaluating} propose the \humaneval dataset and show LLMs can generate functionally correct implementations by introducing a functional correctness evaluation process.
Concurrently, \citet{austin2021program} propose the MBPP dataset for Python basic programs and \citet{hendrycks2021measuring} release the APPS dataset related to coding contest problems.
Consecutive studies have proposed datasets targeted for realistic purposes.
\citet{lai2023ds} focused on data science problems and \citet{wang-etal-2023-execution} paid attention to realistic coding queries from StackOverflow and \citet{codereval} aimed at Python and Java code generation tasks from real-world open-source projects, and \citet{babe2023studenteval} concentrated on beginning programmers.
These work are combined and included in several coding benchmarks \citep{lu2021codexglue,khan2023xcodeeval,ni2023l2ceval}.
For the metrics, \citet{dong2023codescore} propose CodeScore to estimate functional correctness and \citet{zhou-etal-2023-codebertscore} propose CodeBERTScore that utilizes BERTScore \citep{Zhang2020BERTScore:}.

There exists research work that extends the \humaneval dataset to support other features.
\citet{cassano2022multiple,10.1145/3580305.3599790} extend the dataset to support multiple programming languages and \citet{liu2023is} propose the HumanEval+ dataset that extends their test case to enable rigorous evaluation of functional correctness.
\citet{wang-etal-2023-recode} focused on prompt robustness by extending the \humaneval dataset.
However, an evaluation procedure that enables systematic analysis of how LLMs leverage auxiliary functions has yet to be released in code generation tasks.

\section{Dataset}
\label{sec:dataset}
We manually construct a variety of coding examples with corresponding auxiliary functions.
To do this, we treat the Python examples in the \humaneval dataset as our base auxiliary functions and employ human experts to create an extended function for each example.
We guide them to produce functions that have additional functionalities compared to the given functions.
The following aspects are considered to remove the ambiguity inside the concept of extension and enhance their quality.

\paragraph{Extension type}
There exist two different types of extension, i.e., black-box extension and white-box extension.
The black-box extension extends a function by calling the auxiliary function.
It does not consider the internal mechanism of the auxiliary function.
However, the white-box extension extends them by rewriting the improved internal mechanism.
We allow any type of extension, but recommend the black-box one as calling the existing functions if possible is mostly better than rewriting the whole mechanism \citep{fowler2018refactoring}.

\paragraph{Software engineering concept}
We show the Liskov-substitution principle and the concept of subtyping \citep{liskov1994behavioral} to the labelers.
In doing so, we expect that the curated function could be treated as an extended version of the given function from the software engineering point of view.

\paragraph{Quality control}
We filtered out some examples in the \humaneval dataset that are not appropriate for using auxiliary functions.
We removed the examples that provide the same functionality embedded in Python built-in functions, e.g., sum\_to\_n, as it already serves through the Python features.
Also, the examples that are semantically duplicated with other examples are excluded from the final evaluation set.
For example, if the two functions handle the same logic to process symbols but accept brackets or parentheses as their inputs, one of them is removed.

\paragraph{}
We collect 151 problems representing a function pair that one function extends the other and name it \proposeddataset.
Additionally, we mechanically parse these code snippets and create features for components for future usage.

\section{Experiments}
\label{sec:experiments}
We comprehensively evaluate LLMs' ability to harness auxiliary functions using our \proposeddataset dataset.
To do this, we designed research questions as follows.
\begin{itemize}[leftmargin=10pt]
\setlength{\itemsep}{0pt}
\item \textbf{RQ1}: Could LLMs properly and robustly utilize different types of auxiliary functions?
\item \textbf{RQ2}: How do LLMs' implementations vary when they access relevant auxiliary functions?
\item \textbf{RQ3}: Do current training methodologies enhance the ability to utilize auxiliary functions?
\end{itemize}
We first examine the effectiveness and robustness of including a single auxiliary function in the prompt and extend this setting into multiple auxiliary functions.
Also, we explore their implementation styles and analyze them based on human preference.

\subsection{Single auxiliary function experiment}

We measure the effectiveness of an auxiliary function in a code synthesis task by designing several prompts varying their existence and type.
Currently, the prompt used in the existing work to solve the task is mainly composed of a target function signature with the corresponding import statements \citep{bigcode-evaluation-harness,cassano2022multiple,chen2021evaluating}.
We attached the auxiliary function signature and their implementation between the import statements and the target function signature to allow LLMs to access the knowledge about auxiliary functions.
Our prompts with several types of auxiliary functions are described as follows.
\begin{itemize}[leftmargin=10pt]
    \setlength{\itemsep}{0pt}
    \item \textbf{No auxiliary function (Direct)}:
    Prompt consists of a target function signature without auxiliary functions.
    This setting acts as a baseline in our experiments.
    \item \textbf{Human-written irrelevant auxiliary function (Irrelevant)}:
    We attached an irrelevant auxiliary function written by humans in the prompt.
    We constructed an auxiliary function pool with the canonical solutions in the \humaneval dataset \citep{chen2021evaluating} and sampled an irrelevant function from the pool to construct the prompt.
    \item \textbf{Model-written relevant auxiliary function (Step-by-step)}:
    We utilize the relevant auxiliary function written by the model in the prompt.
    Concretely, LLMs first synthesize relevant auxiliary function and then it is attached to the prompt for implementing the target function. 
    Note that only a relevant auxiliary function signature without their implementation is additionally required for this setting.
    We utilize the auxiliary function signatures in the \humaneval dataset and the target one in our \proposeddataset dataset.
    \item \textbf{Human-written relevant auxiliary function (Oracle)}:
    We provide a relevant auxiliary function written by humans to the model.
    The corresponding canonical solutions in the \humaneval dataset are used for human-written relevant auxiliary functions.
    We consider this setting as an oracle because these functions are currently the best in terms of quality and understandability.
\end{itemize}
The details about function signature, e.g., type annotation and docstring format, are consistent with the format curated in \citet{cassano2022multiple}.

\paragraph{Language models}
We collect several LLMs pre-trained on code described as follows.
\begin{itemize}[leftmargin=10pt]
    \setlength{\itemsep}{0pt}
    \item \textbf{Incoder} \citep{fried2023incoder} is the early open-source decoder-only generative language model pretrained on public codes and StackOverflow questions and answers. 
    \item \textbf{CodeGen} \citep{nijkamp2023codegen} is another open-source language model pretrained on public codes.
    We use two versions where “Multi” represents pre-training on multiple programming languages and “Mono” is additionally trained on Python codes from the "Multi" checkpoint.
    \item \textbf{BigCode} \citep{allal2023santacoder,li2023starcoder}  releases two checkpoints, i.e., SantaCoder and StarCoder, pretrained on public codes.
    They adopt various data-cleaning techniques to enhance the quality of the training corpus.
    \item \textbf{CodeLLaMA} \citep{rozière2023code} is a variant of LLaMA2 \citep{touvron2023llama2} additionally pretrained on code corpus.
    CodeLLaMAPython and CodeLLaMAInstruct are further trained on Python codes and instruction following datasets, respectively.
\end{itemize}

\paragraph{Decoding strategy}
We follow the decoding strategy for LLMs consistent with the existing benchmark \citep{bigcode-evaluation-harness}.
We use nucleus sampling \citep{Holtzman2020The} with top-p 0.95 and low-temperature scaling, i.e., 0.2, focusing on the correctness of the generated implementation.
The models generate at most 512 tokens for each prompt and stop generation when either end of sequence token or predefined stop sequences, i.e., "\textbackslash ndef", "\textbackslash nclass", "\textbackslash nif", "\textbackslash n\#", are generated.

\paragraph{Evaluation criteria}
The implementations generated by the models are evaluated on functional correctness based on the corresponding test cases.
Specifically, an implementation is regarded as functionally correct when it passes all the corresponding test cases.
We use the widely used pass@1 metric indicating the proportion of functionally corrected implementations among generated implementations.
To reduce the variance of the pass@1 metric, we generate eight implementations for each problem when estimating the model performance.

\begin{table*}[t]
\centering
\begin{tabular}{@{}lcccc@{}}
\toprule
\multicolumn{1}{c}{Model} & Direct & Irrelevant & Step-by-step & Oracle \\ \midrule
Incoder 1B & 0.0373 & 0.0472 (+26.7\%) & 0.0364 (-2.2\%) & 0.2028 (+444.4\%) \\
Incoder 6B & 0.0621 & 0.0762 (+22.7\%) & 0.0737 (+18.7\%) & 0.2856 (+360.0\%) \\
CodeGenMulti 2B & 0.0969 & 0.0894 (-7.7\%) & 0.0778 (-19.7\%) & 0.2856 (+194.9\%) \\
CodeGenMulti 16B & 0.1060 & 0.1134 (+7.0\%) & 0.1093 (+3.1\%) & 0.3568 (+236.7\%) \\
CodeGenMono 2B & 0.1068 & 0.1118 (+4.7\%) & 0.1366 (+27.9\%) & 0.3469 (+224.8\%) \\
CodeGenMono 16B & 0.1912 & 0.1912 (0.0\%) & 0.2127 (+11.3\%) & 0.4776 (+149.8\%) \\
Santacoder 1B & 0.1002 & 0.1010 (+0.8\%) & 0.0944 (-5.8\%) & 0.3104 (+209.9\%) \\
Starcoder 16B & 0.1937 & 0.2310 (+19.2\%) & 0.2848 (+47.0\%) & 0.5596 (+188.9\%) \\
CodeLLaMA 7B & 0.1738 & 0.2185 (+25.7\%) & 0.2219 (+27.6\%) & 0.5248 (+201.9\%) \\
CodeLLaMA 13B & 0.2359 & 0.2773 (+17.5\%) & 0.2773 (+17.5\%) & 0.5712 (+142.1\%) \\
CodeLLaMA 34B & 0.2748 & 0.3262 (+18.7\%) & 0.3750 (+36.4\%) & 0.6416 (+133.4\%) \\
CodeLLaMAPython 7B & 0.2583 & 0.2690 (+4.2\%) & 0.3237 (+25.3\%) & 0.5919 (+129.2\%) \\
CodeLLaMAPython 13B & 0.2657 & 0.3278 (+23.4\%) & 0.3957 (+48.9\%) & 0.5737 (+115.9\%) \\
CodeLLaMAPython 34B & 0.3179 & 0.3460 (+8.9\%) & 0.4296 (+35.2\%) & 0.6598 (+107.6\%) \\
CodeLLaMAInstruct 7B & 0.2955 & 0.3088 (+4.5\%) & 0.3526 (+19.3\%) & 0.4164 (+40.9\%) \\
CodeLLaMAInstruct 13B & 0.3874 & 0.3791 (-2.1\%) & 0.4172 (+7.7\%) & 0.5017 (+29.5\%) \\
CodeLLaMAInstruct 34B & 0.4222 & 0.4214 (-0.2\%) & 0.4255 (+0.8\%) & 0.5017 (+18.8\%) \\ \bottomrule
\end{tabular}
\caption{The pass@1 performance on the \proposeddataset dataset. The values in the parentheses represent the relative improvement with the Direct setting.}
\label{tab:exp1}
\end{table*}

\subsubsection{Performance analysis}
We report the performances and the relative improvement compared with the one without auxiliary function in Table~\ref{tab:exp1} and compare them to identify the effectiveness of different auxiliary functions.

\paragraph{Effects on human-written relevant auxiliary function}
Whole models exhibit remarkable improvement when they access the human-written relevant auxiliary functions (Table~\ref{tab:exp1}, Oracle).
It implies that most LLMs could utilize the proper relevant auxiliary function.
The improvement is observed even for the most recent competitive model, i.e., CodeLLaMAPython 34B, indicating assisting code synthesis with auxiliary function is still a valid approach even as the model size grows.

\begin{figure*}
\centering
\begin{subfigure}[b]{0.48\textwidth}
\centering
\includegraphics[width=\textwidth]{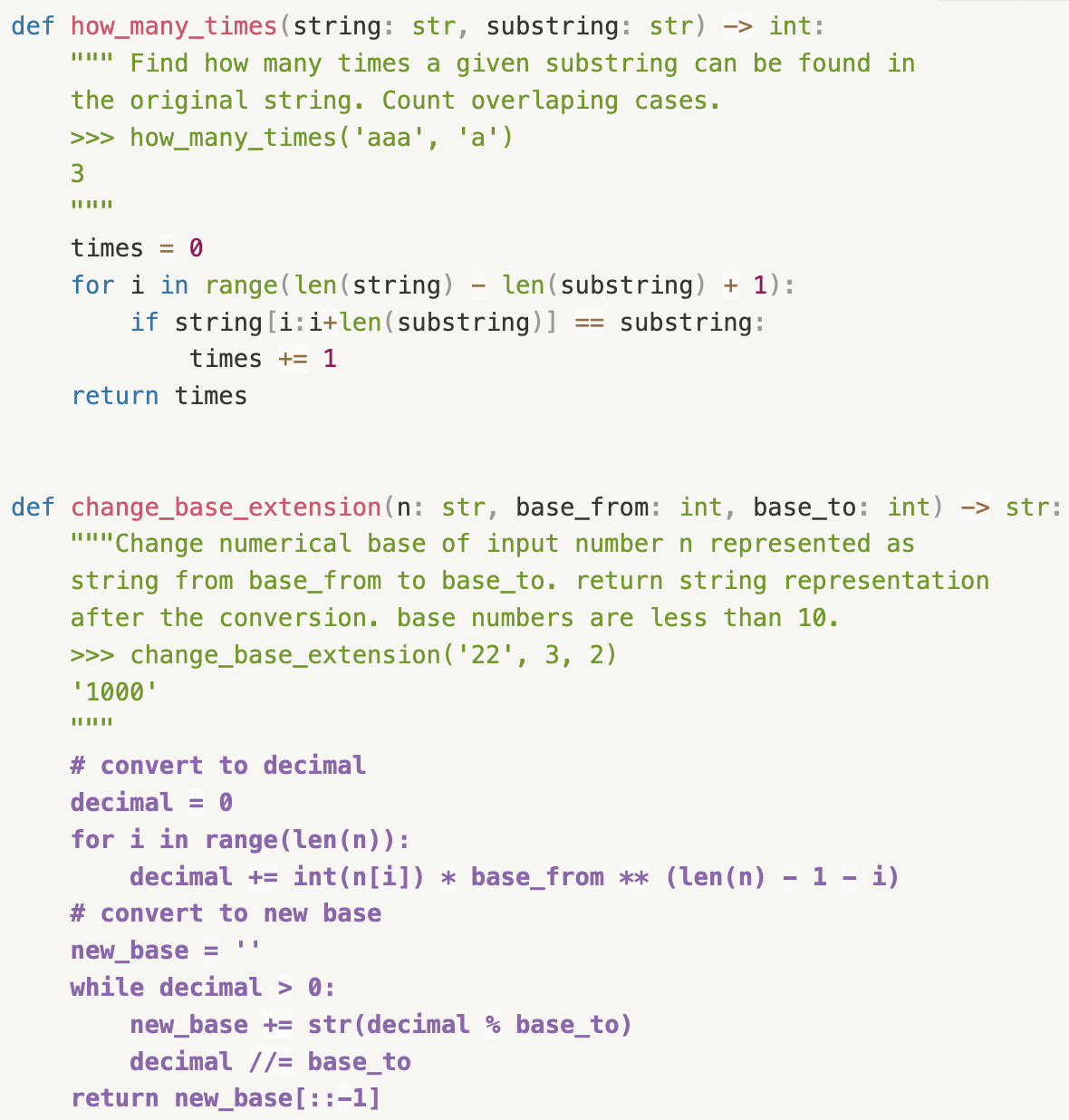}
\caption{A passed case with irrelevant auxiliary function.}
\label{fig:example_irrelevant}
\end{subfigure}
\hfill
\begin{subfigure}[b]{0.48\textwidth}
\centering
\includegraphics[width=\textwidth]{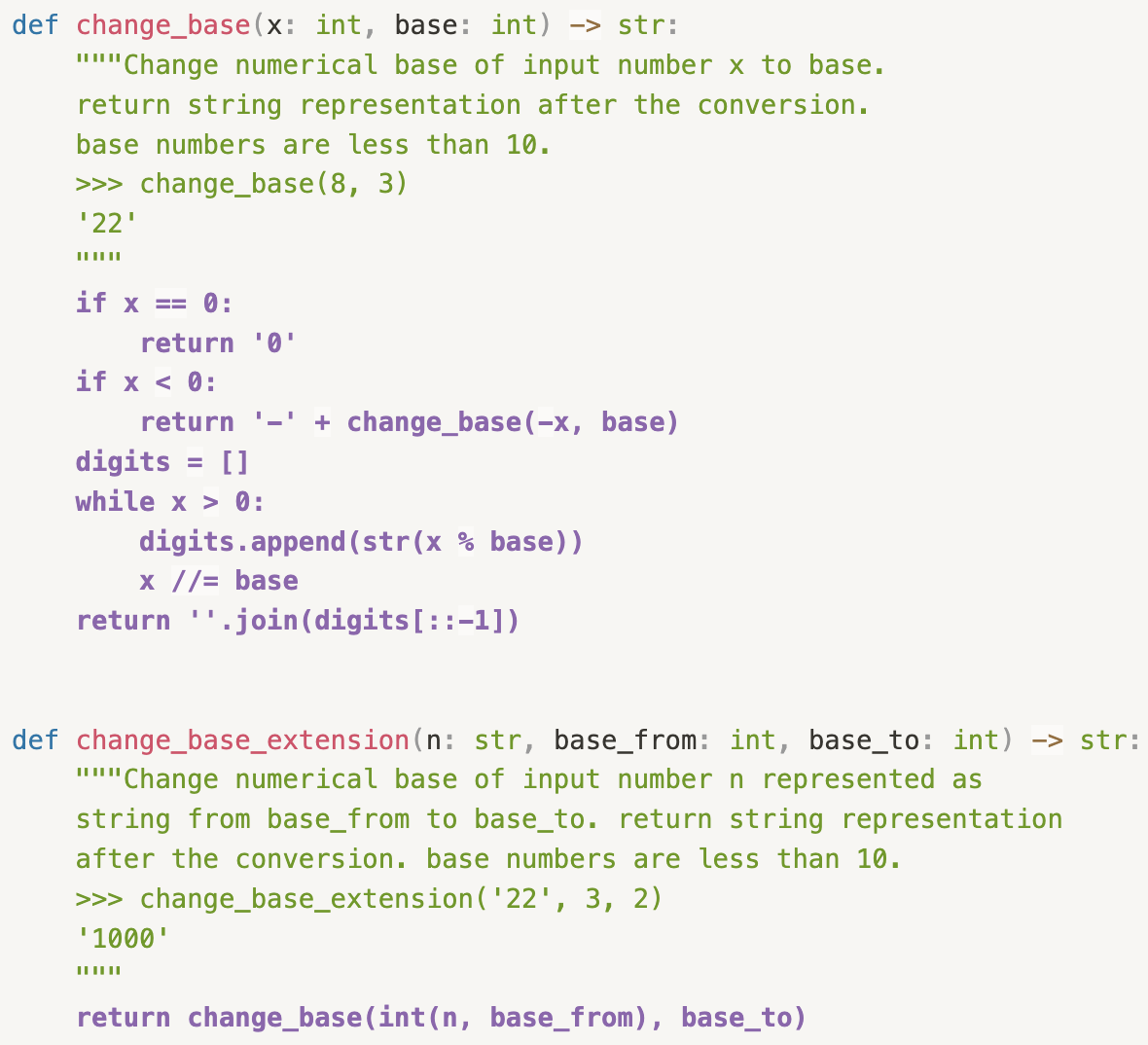}
\caption{A passed case with relevant auxiliary function.}
\label{fig:example_relevant}
\end{subfigure}
\caption{The two types of generated implementations from CodeLLaMAPython 13B. Bold purple texts are generated by the model while the others are given. Some examples in the docstring are omitted for brevity.}
\label{fig:case_study}
\end{figure*}

\paragraph{Effects on model-written relevant auxiliary function}
Considering the "Step-by-step" column in Table~\ref{tab:exp1}, the model-written relevant auxiliary functions contribute to the improvement for some advanced LLMs.
CodeLLaMA series, StarCoder, CodeGenMono series, and Incoder 6B properly utilize the auxiliary function written by themselves.
It suggests that the models can improve their codes if we provide a two-step plan in the form of function signatures.
We attach one successful example that calls the generated auxiliary function during target implementation in Figure~\ref{fig:example_relevant}.
In this sense, this approach is similar to the Least-to-Most prompting \citep{zhou2023leasttomost} that solves target tasks with the model-generated answer of predefined subtasks.

\paragraph{Effects on human-written irrelevant auxiliary function}
We observe that providing an irrelevant auxiliary function brings meaningful improvement on few models.
To investigate how these functions affect the target implementation, we qualitatively analyze the examples that CodeLLaMAPython 13B successfully generates under both settings, i.e., irrelevant and step-by-step.
In Figure~\ref{fig:case_study}, we found that the irrelevant auxiliary function acts as a demonstration like few-shot prompting so that the few models exhibit performance improvement.
However, since the given auxiliary function is not relevant to the target function (Figure~\ref{fig:example_irrelevant}), no implementation pattern that directly utilizes the auxiliary function is found.
On the contrary, the relevant auxiliary functions are successfully utilized by calling in the target function and reduce their implementation difficulty (Figure~\ref{fig:example_relevant}).
Therefore, we conclude there exists a unique advantage of providing relevant auxiliary function although the irrelevant one is helpful to some extent.

\paragraph{Effects on Python specialization}
We investigate how the additional training with Python corpus affects its ability to utilize auxiliary functions.
To do this, we compare the two model families specialized in Python, i.e., CodeGenMono and CodeLLaMAPython.
In these model groups, we observed higher pass rates compared to the corresponding base model groups, i.e., CodeGenMulti and CodeLLaMA.
Comparing CodeGenMono 2B and CodeGenMulti 2B, the pass rate is similar when no auxiliary function is provided (Direct), but the pass rate of CodeGenMono becomes significantly higher than that of CodeGenMulti when we provide an appropriate auxiliary function (Oracle).
Additionally, in the Step-by-step setting, CodeGenMono models show meaningful improvement while CodeGenMulti could not.
In the case of CodeLLaMA, CodeLLaMAPython models show higher pass rates in the whole model size.
From these experimental evidences, we conclude that additional learning with Python code enhances the ability to utilize auxiliary functions.
We speculate that the Python codes used for training contain relevant functions in the same file and the model is trained to jointly consider the functions within the same context.

\paragraph{Effects on instruction tuning}
We also compare CodeLLaMAInstruct models to determine whether the instruction tuning affects the ability to harness auxiliary functions.
In order to use an instruction-tuned model, instructions written in natural language and a prompt template are additionally required.
To this end, we apply an approach similar to HumanEvalPack~\citep{muennighoff2023octopack}, where the instructions are automatically generated from the original prompt.
We combine these instructions with the CodeLLaMAInstruct template to create a prompt.
The prompt is formulated into two consecutive turns where the first turn is about the auxiliary function and the second one is about generating the target function\footnote{Refer to the appendix for the detailed template structure.}.

Our empirical results show that CodeLLaMAInstruct models perform better than CodeLLaMA models when implementing functions without auxiliary functions (Table~\ref{tab:exp1}, Direct), which is consistent with previous findings \citep{rozière2023code}.
On the other hand, when an appropriate auxiliary function is provided in the prompt (Table~\ref{tab:exp1}, Oracle), the base models show better performance than the instruction-tuned models.
In addition, the relative improvement in the Step-by-step settings has prominently decreased compared to that of the base models.
This suggests that the ability to utilize other functions in the context has been weakened during the instruction tuning process.
Therefore, it is necessary to develop an advanced instruction-tuning methodology to incorporate the previously implemented functions, which is our future work.

\begin{figure}
\centering
\includegraphics[width=\columnwidth]{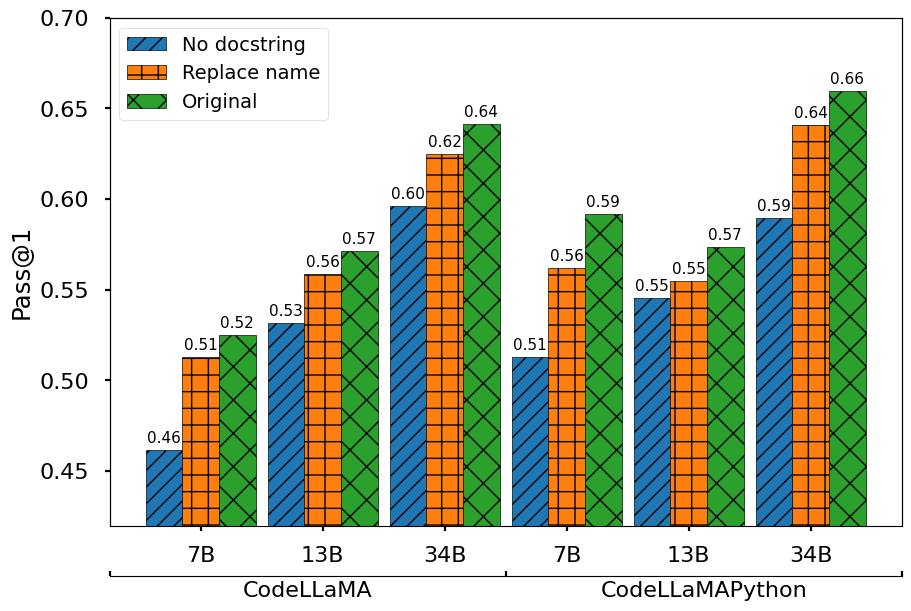}
\caption{Robustness analysis on two perturbations, renaming auxiliary function and removing docstring.}
\label{fig:ablation}
\end{figure}

\subsubsection{Robustness analysis}

We check whether the model could properly use the given relevant auxiliary function after some components inside the function have been perturbed.
We apply two perturbations: (1) replacing the name of the auxiliary function with other function names in the \humaneval dataset or (2) deleting the docstring included in the function.
Note that the functionality of the auxiliary function itself does not change because we did not change the function implementation or its input/output format.

\paragraph{Results}
The experimental results show that even if the functionality of the function does not change, a performance drop is observed depending on the name of the function or the existence of a docstring (Figure~\ref{fig:ablation}).
The lack of a docstring had a greater impact than renaming the function, and it is natural in that the docstring contains a more detailed description of its functionality.
Despite their usefulness, we want to highlight that LLMs have to understand the function without docstring for their realistic use cases as most practical codes do not include them.\footnote{In \texttt{bigcode/the-stack-smol}, 70.5\% of Python functions do not have docstring.}
The performance drop was not alleviated even when the model size was increased or the model was additionally trained with Python codes.
Therefore, there is a need to propose a robust learning methodology that can reduce performance differences caused by such perturbations.

\subsubsection{Implementation style analysis}

We analyze the generated implementation based on their style and compare preferences between them.
In this experiment, we use the implementations generated under the Oracle setting.
To identify their implementation style, we apply Python static parser\footnote{\url{https://docs.python.org/3/library/ast}} and check whether they called the given auxiliary function.
The implementations that call the auxiliary function are regarded as black-box style while the rest as white-box style.
The black-box style directly utilizes the auxiliary function as is, while the white-box style mimics the internal mechanism of the auxiliary function.

\begin{figure}
\centering
\includegraphics[width=\columnwidth]{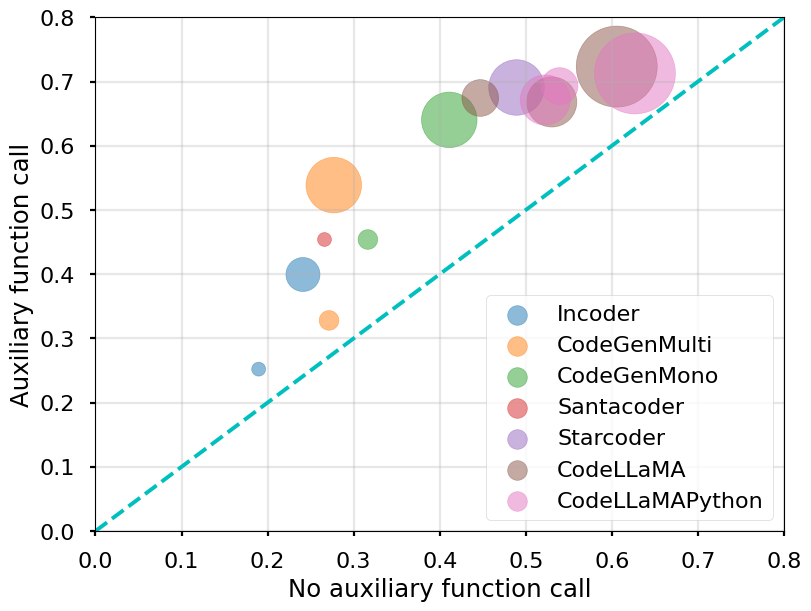}
\caption{Pass@1 score comparison between black-box implementations (auxiliary function call) and white-box implementations (no auxiliary function call). The scale of dots represents their model size. Aqua dotted line indicates the performance on black-box and white-box implementations are the same.}
\label{fig:aux_func_call_analysis}
\end{figure}

\paragraph{Results}
We compute pass@1 scores for each style and model (Figure~\ref{fig:aux_func_call_analysis}).
The results show that all models can implement functions in both styles.
Also, we observed that the pass@1 score for the black-box style is higher than that of the white-box style.
It implies that calling an auxiliary function is much safer and more accurate if the target function can be implemented by calling the auxiliary function.
Currently, up to 40\% of the model-generated implementations are implemented in black-box style, even though most examples can be implemented in black-box style.\footnote{147 of 151 human-written reference solutions in the \proposeddataset dataset are black-box style.}
Therefore, it is expected that the pass@1 score can be improved if more examples are implemented in black box style.
Furthermore, we would like to emphasize that the improvement of the ability to generate black-box implementations is diminishing as language models evolve.
This phenomenon suggests that model developers should consider the model's function call ability when learning their models.

\paragraph{Human evaluation}
Further investigating the two different styles, we conduct a human pairwise preference evaluation with human-written implementations (Human), and model-written ones with both styles (Black box and White box).
We created a labeling sheet with 17 examples that CodeLLaMAPython 34B implements in both styles correctly.
We recruited labelers who have been coding with Python for over five years.
For the three possible pairs, labelers were instructed to choose the better implementations according to their preference such as performance or readability.

The evaluation results in Table~\ref{tab:human_eval} show that implementations that call auxiliary functions are preferred over implementations that do not.
After inspecting the result qualitatively, we interpret that most black box implementations were selected due to their clarity and conciseness coming from appropriately delegating subroutines to auxiliary functions.
Usually, the model-generated white-box implementations tend to repeat the identical mechanism inside the auxiliary function, which is not preferred in software engineering fields \citep{10.5555/320326}.
In few cases, white-box implementations are preferred over black-box ones as they are considered as over-engineering.
Therefore, training the models to delegate the subroutine to other functions suitably would be the next step for generating realistic code.

\begin{table}[]
\centering
\resizebox{\columnwidth}{!}{%
\begin{tabular}{@{}lccc@{}}
\toprule
\multicolumn{1}{c}{} & Former Win & Tie & Latter Win \\ \midrule
Human vs Black box & 37.25 & 56.86 & 5.88 \\
Human vs White box & 88.24 & 1.96 & 9.80 \\
Black box vs White box & 84.31 & 5.88 & 9.80 \\ \bottomrule
\end{tabular}%
}
\caption{Human pairwise preference evaluation results}
\label{tab:human_eval}
\end{table}

\begin{figure}[t]
\centering
\begin{subfigure}[b]{0.48\textwidth}
\includegraphics[width=\textwidth]{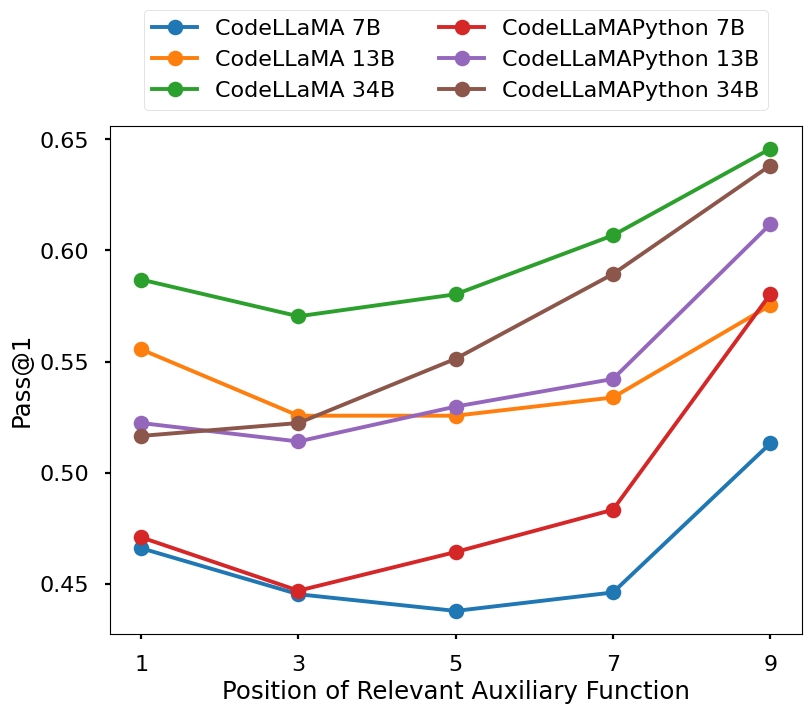}
\caption{Pass@1 scores depending on the position of relevant auxiliary function.}
\label{fig:position_relevant}
\end{subfigure}
% \hfill
\begin{subfigure}[b]{0.48\textwidth}
\includegraphics[width=\textwidth]{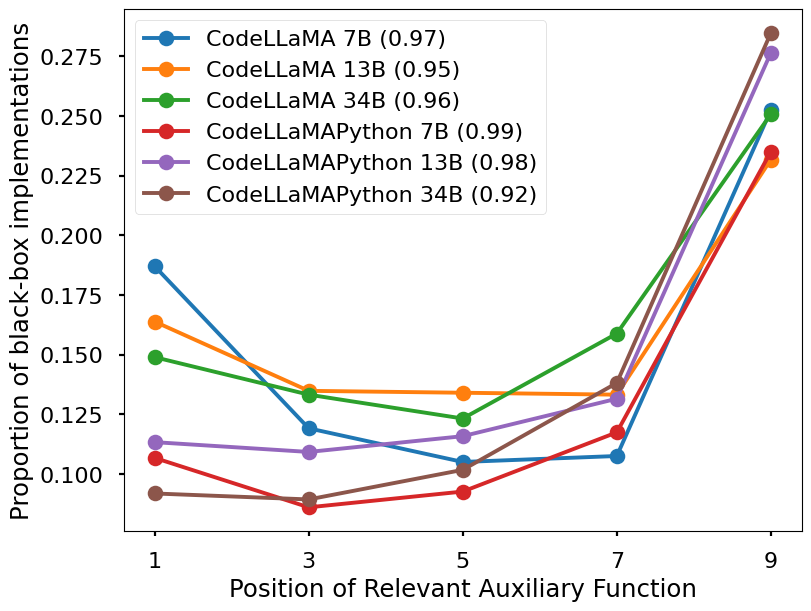}
\caption{Proportion of black-box style implementations among generated ones. The scores inside the parentheses are Pearson correlation scores between the proportion of black-box style implementations and the pass@1 scores.}
\label{fig:prop_black_box}
\end{subfigure}
\caption{Robustness analysis results with multiple auxiliary functions.}
\label{fig:multi_aux_functions}
\end{figure}

\subsection{Multiple auxiliary function experiment}
We provide several auxiliary functions in the prompt and study whether the model selectively utilizes the appropriate auxiliary function.

\subsubsection{Experimental setup}
We design a prompt with nine auxiliary functions followed by the target function signature.
The functions consist of one relevant auxiliary function and the others are randomly sampled from the auxiliary function pool used in the Irrelevant setting.
We change the location of the relevant function in the prompt and measure the pass@1 score and the proportion of black-box style implementations classified as the existence of auxiliary function call.

\subsubsection{Result}
The experimental results are shown in Figure~\ref{fig:multi_aux_functions} and we list up empirical findings we observed.

\paragraph{Performance trends}
We confirmed that CodeLLaMA models and CodeLLaMAPython models show different trends in terms of pass@1 scores (Figure~\ref{fig:position_relevant}).
For CodeLLaMA models, the pass@1 scores showed a U-shape trend, indicating that the performance improved when the related function was located at the first or the last.
This result is consistent with the existing findings \citep{liu2023lost} that, in natural language processing tasks, LLMs can effectively utilize relevant documents when they are located at the beginning or end.
On the other hand, for CodeLLaMAPython models, this U-shape trend was weakened and the pass@1 score increased only when the relevant function was located at the end.
We conjectured that the two related functions were usually located adjacently in Python codes and this pattern was learned by the model.
However, since the location of relevant functions is independent of their functionality, LLMs need to be tuned to robustly utilize them regardless of where they are placed.

\paragraph{Correlation with black-box style implementations}
We found that there exists a strong correlation between the pass@1 score and the proportion of black-box style implementations.
The Pearson correlation scores between the proportion and the pass rate (Figure~\ref{fig:prop_black_box}) are larger than 0.9, indicating that LLMs get higher scores when they try to call appropriate auxiliary functions.
However, the black-box style implementations are mostly observed when the relevant auxiliary functions are located at the last, which provides an explanation of why the pass@1 score is higher when the relevant function is located at the last.
For CodeLLaMA models, they can call the relevant function if they are located at the first, which causes the U-shape trend in pass@1 scores.
Model scaling and additional training with Python codes provide a marginal effect on promoting a model to generate black-box style implementations, suggesting that specialized training for LLMs to call relevant functions similar to invoking general LLMs to use tools \citep{schick2023toolformer} is needed for enhancing their code synthesis ability.

% retrieval augmented generation \citep{parvez-etal-2021-retrieval-augmented,lu-etal-2022-reacc,shrivastava2023repositorylevel}.

\section{Conclusion}
\label{sec:conclusion}
We have explored the ability to utilize auxiliary functions encoded in the LLMs through our newly proposed \proposeddataset dataset.
The \proposeddataset dataset is constructed to contain function relationships while considering the software engineering concepts.
Our multi-faceted experiments with the \proposeddataset dataset comprehensively show the current LLMs' ability to harness auxiliary functions.
Our auxiliary function experiments demonstrate that the LLMs have the ability to utilize auxiliary functions even when the function is implemented by themselves.
However, our in-depth analysis discovered that their ability varies depending on the factors that humans might not, i.e., the position of relevant functions, function name, and docstring.
Furthermore, our implementation style analysis reveals that, in some cases, the LLMs repeat the mechanism of the given auxiliary function while humans simply call the auxiliary functions, suggesting the future research direction of current code LLMs for auxiliary function calls.

\section{Limitations}
\label{sec:limitations}
Although the curated dataset in this study allowed us to evaluate the ability to utilize auxiliary functions from a variety of perspectives, it has some limitations in determining whether multiple relevant auxiliary functions can be jointly utilized.
Additionally, our behavioral analyses indicate that the capabilities have been empirically observed, but it might be insufficient to conclude the model truly understands and utilizes the auxiliary function, so additional methods are required to reinforce the statement.
% For example, advancing step-by-step generation by automatically planing the function signatures or applying generation more than two steps to pushing the limit of their capability is an interesting future work.

\section*{Acknowledgements}

This work was supported by the Institute of Information \& communications Technology Planning \& Evaluation (IITP) grant funded by the Korea government (MSIT) (No.2018-0-00584, (SW starlab) Development of Decision Support System Software based on Next-Generation Machine Learning and No.2019-0-01906, Artificial Intelligence Graduate School Program (POSTECH)), the National Research Foundation of Korea (NRF) grant funded by the MSIT (No. RS-2023-00217286), and the Digital Innovation Hub project supervised by the Daegu Digital Promotion Agency (DIP) grant funded by the Korean government (MSIT and Daegu Metropolitan City) in 2024 (No. DBSD1-07).

\bibliography{anthology,custom}

\begin{thebibliography}{43}
\expandafter\ifx\csname natexlab\endcsname\relax\def\natexlab#1{#1}\fi

\bibitem[{Allal et~al.(2023)Allal, Li, Kocetkov, Mou, Akiki, Ferrandis, Muennighoff, Mishra, Gu, Dey, Umapathi, Anderson, Zi, Poirier, Schoelkopf, Troshin, Abulkhanov, Romero, Lappert, Toni, del Río, Liu, Bose, Bhattacharyya, Zhuo, Yu, Villegas, Zocca, Mangrulkar, Lansky, Nguyen, Contractor, Villa, Li, Bahdanau, Jernite, Hughes, Fried, Guha, de~Vries, and von Werra}]{allal2023santacoder}
Loubna~Ben Allal, Raymond Li, Denis Kocetkov, Chenghao Mou, Christopher Akiki, Carlos~Munoz Ferrandis, Niklas Muennighoff, Mayank Mishra, Alex Gu, Manan Dey, Logesh~Kumar Umapathi, Carolyn~Jane Anderson, Yangtian Zi, Joel~Lamy Poirier, Hailey Schoelkopf, Sergey Troshin, Dmitry Abulkhanov, Manuel Romero, Michael Lappert, Francesco~De Toni, Bernardo~García del Río, Qian Liu, Shamik Bose, Urvashi Bhattacharyya, Terry~Yue Zhuo, Ian Yu, Paulo Villegas, Marco Zocca, Sourab Mangrulkar, David Lansky, Huu Nguyen, Danish Contractor, Luis Villa, Jia Li, Dzmitry Bahdanau, Yacine Jernite, Sean Hughes, Daniel Fried, Arjun Guha, Harm de~Vries, and Leandro von Werra. 2023.
\newblock \href {http://arxiv.org/abs/2301.03988} {Santacoder: don't reach for the stars!}

\bibitem[{Austin et~al.(2021)Austin, Odena, Nye, Bosma, Michalewski, Dohan, Jiang, Cai, Terry, Le, and Sutton}]{austin2021program}
Jacob Austin, Augustus Odena, Maxwell Nye, Maarten Bosma, Henryk Michalewski, David Dohan, Ellen Jiang, Carrie Cai, Michael Terry, Quoc Le, and Charles Sutton. 2021.
\newblock \href {http://arxiv.org/abs/2108.07732} {Program synthesis with large language models}.

\bibitem[{Babe et~al.(2023)Babe, Nguyen, Zi, Guha, Feldman, and Anderson}]{babe2023studenteval}
Hannah~McLean Babe, Sydney Nguyen, Yangtian Zi, Arjun Guha, Molly~Q Feldman, and Carolyn~Jane Anderson. 2023.
\newblock \href {http://arxiv.org/abs/2306.04556} {Studenteval: A benchmark of student-written prompts for large language models of code}.

\bibitem[{Ben~Allal et~al.(2022)Ben~Allal, Muennighoff, Kumar~Umapathi, Lipkin, and von Werra}]{bigcode-evaluation-harness}
Loubna Ben~Allal, Niklas Muennighoff, Logesh Kumar~Umapathi, Ben Lipkin, and Leandro von Werra. 2022.
\newblock A framework for the evaluation of code generation models.
\newblock \url{https://github.com/bigcode-project/bigcode-evaluation-harness}.

\bibitem[{Cassano et~al.(2022)Cassano, Gouwar, Nguyen, Nguyen, Phipps-Costin, Pinckney, Yee, Zi, Anderson, Feldman, Guha, Greenberg, and Jangda}]{cassano2022multiple}
Federico Cassano, John Gouwar, Daniel Nguyen, Sydney Nguyen, Luna Phipps-Costin, Donald Pinckney, Ming-Ho Yee, Yangtian Zi, Carolyn~Jane Anderson, Molly~Q Feldman, Arjun Guha, Michael Greenberg, and Abhinav Jangda. 2022.
\newblock \href {http://arxiv.org/abs/2208.08227} {Multipl-e: A scalable and extensible approach to benchmarking neural code generation}.

\bibitem[{Chen et~al.(2021)Chen, Tworek, Jun, Yuan, de~Oliveira~Pinto, Kaplan, Edwards, Burda, Joseph, Brockman, Ray, Puri, Krueger, Petrov, Khlaaf, Sastry, Mishkin, Chan, Gray, Ryder, Pavlov, Power, Kaiser, Bavarian, Winter, Tillet, Such, Cummings, Plappert, Chantzis, Barnes, Herbert-Voss, Guss, Nichol, Paino, Tezak, Tang, Babuschkin, Balaji, Jain, Saunders, Hesse, Carr, Leike, Achiam, Misra, Morikawa, Radford, Knight, Brundage, Murati, Mayer, Welinder, McGrew, Amodei, McCandlish, Sutskever, and Zaremba}]{chen2021evaluating}
Mark Chen, Jerry Tworek, Heewoo Jun, Qiming Yuan, Henrique~Ponde de~Oliveira~Pinto, Jared Kaplan, Harri Edwards, Yuri Burda, Nicholas Joseph, Greg Brockman, Alex Ray, Raul Puri, Gretchen Krueger, Michael Petrov, Heidy Khlaaf, Girish Sastry, Pamela Mishkin, Brooke Chan, Scott Gray, Nick Ryder, Mikhail Pavlov, Alethea Power, Lukasz Kaiser, Mohammad Bavarian, Clemens Winter, Philippe Tillet, Felipe~Petroski Such, Dave Cummings, Matthias Plappert, Fotios Chantzis, Elizabeth Barnes, Ariel Herbert-Voss, William~Hebgen Guss, Alex Nichol, Alex Paino, Nikolas Tezak, Jie Tang, Igor Babuschkin, Suchir Balaji, Shantanu Jain, William Saunders, Christopher Hesse, Andrew~N. Carr, Jan Leike, Josh Achiam, Vedant Misra, Evan Morikawa, Alec Radford, Matthew Knight, Miles Brundage, Mira Murati, Katie Mayer, Peter Welinder, Bob McGrew, Dario Amodei, Sam McCandlish, Ilya Sutskever, and Wojciech Zaremba. 2021.
\newblock \href {http://arxiv.org/abs/2107.03374} {Evaluating large language models trained on code}.

\bibitem[{Ding et~al.(2023)Ding, Wang, Ahmad, Ramanathan, Nallapati, Bhatia, Roth, and Xiang}]{ding2023cocomic}
Yangruibo Ding, Zijian Wang, Wasi~Uddin Ahmad, Murali~Krishna Ramanathan, Ramesh Nallapati, Parminder Bhatia, Dan Roth, and Bing Xiang. 2023.
\newblock \href {http://arxiv.org/abs/2212.10007} {Cocomic: Code completion by jointly modeling in-file and cross-file context}.

\bibitem[{Dong et~al.(2023)Dong, Ding, Jiang, Li, Li, and Jin}]{dong2023codescore}
Yihong Dong, Jiazheng Ding, Xue Jiang, Zhuo Li, Ge~Li, and Zhi Jin. 2023.
\newblock \href {http://arxiv.org/abs/2301.09043} {Codescore: Evaluating code generation by learning code execution}.

\bibitem[{Fowler(2018)}]{fowler2018refactoring}
Martin Fowler. 2018.
\newblock \emph{Refactoring}.
\newblock Addison-Wesley Professional.

\bibitem[{Fried et~al.(2023)Fried, Aghajanyan, Lin, Wang, Wallace, Shi, Zhong, Yih, Zettlemoyer, and Lewis}]{fried2023incoder}
Daniel Fried, Armen Aghajanyan, Jessy Lin, Sida Wang, Eric Wallace, Freda Shi, Ruiqi Zhong, Scott Yih, Luke Zettlemoyer, and Mike Lewis. 2023.
\newblock \href {https://openreview.net/forum?id=hQwb-lbM6EL} {Incoder: A generative model for code infilling and synthesis}.
\newblock In \emph{The Eleventh International Conference on Learning Representations}.

\bibitem[{Gao et~al.(2023)Gao, Madaan, Zhou, Alon, Liu, Yang, Callan, and Neubig}]{gao2023pal}
Luyu Gao, Aman Madaan, Shuyan Zhou, Uri Alon, Pengfei Liu, Yiming Yang, Jamie Callan, and Graham Neubig. 2023.
\newblock \href {http://arxiv.org/abs/2211.10435} {Pal: Program-aided language models}.

\bibitem[{Gunasekar et~al.(2023)Gunasekar, Zhang, Aneja, Mendes, Giorno, Gopi, Javaheripi, Kauffmann, de~Rosa, Saarikivi, Salim, Shah, Behl, Wang, Bubeck, Eldan, Kalai, Lee, and Li}]{gunasekar2023textbooks}
Suriya Gunasekar, Yi~Zhang, Jyoti Aneja, Caio César~Teodoro Mendes, Allie~Del Giorno, Sivakanth Gopi, Mojan Javaheripi, Piero Kauffmann, Gustavo de~Rosa, Olli Saarikivi, Adil Salim, Shital Shah, Harkirat~Singh Behl, Xin Wang, Sébastien Bubeck, Ronen Eldan, Adam~Tauman Kalai, Yin~Tat Lee, and Yuanzhi Li. 2023.
\newblock \href {http://arxiv.org/abs/2306.11644} {Textbooks are all you need}.

\bibitem[{Hendrycks et~al.(2021)Hendrycks, Basart, Kadavath, Mazeika, Arora, Guo, Burns, Puranik, He, Song, and Steinhardt}]{hendrycks2021measuring}
Dan Hendrycks, Steven Basart, Saurav Kadavath, Mantas Mazeika, Akul Arora, Ethan Guo, Collin Burns, Samir Puranik, Horace He, Dawn Song, and Jacob Steinhardt. 2021.
\newblock \href {https://openreview.net/forum?id=sD93GOzH3i5} {Measuring coding challenge competence with {APPS}}.
\newblock In \emph{Thirty-fifth Conference on Neural Information Processing Systems Datasets and Benchmarks Track (Round 2)}.

\bibitem[{Holtzman et~al.(2020)Holtzman, Buys, Du, Forbes, and Choi}]{Holtzman2020The}
Ari Holtzman, Jan Buys, Li~Du, Maxwell Forbes, and Yejin Choi. 2020.
\newblock \href {https://openreview.net/forum?id=rygGQyrFvH} {The curious case of neural text degeneration}.
\newblock In \emph{International Conference on Learning Representations}.

\bibitem[{Hunt and Thomas(2000)}]{10.5555/320326}
Andrew Hunt and David Thomas. 2000.
\newblock \emph{The Pragmatic Programmer: From Journeyman to Master}.
\newblock Addison-Wesley Longman Publishing Co., Inc., USA.

\bibitem[{Iyer et~al.(2018)Iyer, Konstas, Cheung, and Zettlemoyer}]{iyer-etal-2018-mapping}
Srinivasan Iyer, Ioannis Konstas, Alvin Cheung, and Luke Zettlemoyer. 2018.
\newblock \href {https://doi.org/10.18653/v1/D18-1192} {Mapping language to code in programmatic context}.
\newblock In \emph{Proceedings of the 2018 Conference on Empirical Methods in Natural Language Processing}, pages 1643--1652, Brussels, Belgium. Association for Computational Linguistics.

\bibitem[{Khan et~al.(2023)Khan, Bari, Do, Wang, Parvez, and Joty}]{khan2023xcodeeval}
Mohammad Abdullah~Matin Khan, M~Saiful Bari, Xuan~Long Do, Weishi Wang, Md~Rizwan Parvez, and Shafiq Joty. 2023.
\newblock \href {http://arxiv.org/abs/2303.03004} {xcodeeval: A large scale multilingual multitask benchmark for code understanding, generation, translation and retrieval}.

\bibitem[{Lai et~al.(2023)Lai, Li, Wang, Zhang, Zhong, Zettlemoyer, Yih, Fried, Wang, and Yu}]{lai2023ds}
Yuhang Lai, Chengxi Li, Yiming Wang, Tianyi Zhang, Ruiqi Zhong, Luke Zettlemoyer, Wen-tau Yih, Daniel Fried, Sida Wang, and Tao Yu. 2023.
\newblock Ds-1000: A natural and reliable benchmark for data science code generation.
\newblock In \emph{International Conference on Machine Learning}, pages 18319--18345. PMLR.

\bibitem[{Li et~al.(2023)Li, Allal, Zi, Muennighoff, Kocetkov, Mou, Marone, Akiki, Li, Chim, Liu, Zheltonozhskii, Zhuo, Wang, Dehaene, Davaadorj, Lamy-Poirier, Monteiro, Shliazhko, Gontier, Meade, Zebaze, Yee, Umapathi, Zhu, Lipkin, Oblokulov, Wang, Murthy, Stillerman, Patel, Abulkhanov, Zocca, Dey, Zhang, Fahmy, Bhattacharyya, Yu, Singh, Luccioni, Villegas, Kunakov, Zhdanov, Romero, Lee, Timor, Ding, Schlesinger, Schoelkopf, Ebert, Dao, Mishra, Gu, Robinson, Anderson, Dolan-Gavitt, Contractor, Reddy, Fried, Bahdanau, Jernite, Ferrandis, Hughes, Wolf, Guha, von Werra, and de~Vries}]{li2023starcoder}
Raymond Li, Loubna~Ben Allal, Yangtian Zi, Niklas Muennighoff, Denis Kocetkov, Chenghao Mou, Marc Marone, Christopher Akiki, Jia Li, Jenny Chim, Qian Liu, Evgenii Zheltonozhskii, Terry~Yue Zhuo, Thomas Wang, Olivier Dehaene, Mishig Davaadorj, Joel Lamy-Poirier, João Monteiro, Oleh Shliazhko, Nicolas Gontier, Nicholas Meade, Armel Zebaze, Ming-Ho Yee, Logesh~Kumar Umapathi, Jian Zhu, Benjamin Lipkin, Muhtasham Oblokulov, Zhiruo Wang, Rudra Murthy, Jason Stillerman, Siva~Sankalp Patel, Dmitry Abulkhanov, Marco Zocca, Manan Dey, Zhihan Zhang, Nour Fahmy, Urvashi Bhattacharyya, Wenhao Yu, Swayam Singh, Sasha Luccioni, Paulo Villegas, Maxim Kunakov, Fedor Zhdanov, Manuel Romero, Tony Lee, Nadav Timor, Jennifer Ding, Claire Schlesinger, Hailey Schoelkopf, Jan Ebert, Tri Dao, Mayank Mishra, Alex Gu, Jennifer Robinson, Carolyn~Jane Anderson, Brendan Dolan-Gavitt, Danish Contractor, Siva Reddy, Daniel Fried, Dzmitry Bahdanau, Yacine Jernite, Carlos~Muñoz Ferrandis, Sean Hughes, Thomas Wolf, Arjun Guha, Leandro von
  Werra, and Harm de~Vries. 2023.
\newblock \href {http://arxiv.org/abs/2305.06161} {Starcoder: may the source be with you!}

\bibitem[{Li et~al.(2022)Li, Choi, Chung, Kushman, Schrittwieser, Leblond, Eccles, Keeling, Gimeno, Lago, Hubert, Choy, de~Masson~d’Autume, Babuschkin, Chen, Huang, Welbl, Gowal, Cherepanov, Molloy, Mankowitz, Robson, Kohli, de~Freitas, Kavukcuoglu, and Vinyals}]{doi:10.1126/science.abq1158}
Yujia Li, David Choi, Junyoung Chung, Nate Kushman, Julian Schrittwieser, Rémi Leblond, Tom Eccles, James Keeling, Felix Gimeno, Agustin~Dal Lago, Thomas Hubert, Peter Choy, Cyprien de~Masson~d’Autume, Igor Babuschkin, Xinyun Chen, Po-Sen Huang, Johannes Welbl, Sven Gowal, Alexey Cherepanov, James Molloy, Daniel~J. Mankowitz, Esme~Sutherland Robson, Pushmeet Kohli, Nando de~Freitas, Koray Kavukcuoglu, and Oriol Vinyals. 2022.
\newblock \href {https://doi.org/10.1126/science.abq1158} {Competition-level code generation with alphacode}.
\newblock \emph{Science}, 378(6624):1092--1097.

\bibitem[{Liskov and Wing(1994)}]{liskov1994behavioral}
Barbara~H Liskov and Jeannette~M Wing. 1994.
\newblock A behavioral notion of subtyping.
\newblock \emph{ACM Transactions on Programming Languages and Systems (TOPLAS)}, 16(6):1811--1841.

\bibitem[{Liu et~al.(2023{\natexlab{a}})Liu, Xia, Wang, and ZHANG}]{liu2023is}
Jiawei Liu, Chunqiu~Steven Xia, Yuyao Wang, and LINGMING ZHANG. 2023{\natexlab{a}}.
\newblock \href {https://openreview.net/forum?id=1qvx610Cu7} {Is your code generated by chat{GPT} really correct? rigorous evaluation of large language models for code generation}.
\newblock In \emph{Thirty-seventh Conference on Neural Information Processing Systems}.

\bibitem[{Liu et~al.(2023{\natexlab{b}})Liu, Lin, Hewitt, Paranjape, Bevilacqua, Petroni, and Liang}]{liu2023lost}
Nelson~F. Liu, Kevin Lin, John Hewitt, Ashwin Paranjape, Michele Bevilacqua, Fabio Petroni, and Percy Liang. 2023{\natexlab{b}}.
\newblock \href {http://arxiv.org/abs/2307.03172} {Lost in the middle: How language models use long contexts}.

\bibitem[{Lu et~al.(2021)Lu, Guo, Ren, Huang, Svyatkovskiy, Blanco, Clement, Drain, Jiang, Tang, Li, Zhou, Shou, Zhou, Tufano, GONG, Zhou, Duan, Sundaresan, Deng, Fu, and LIU}]{lu2021codexglue}
Shuai Lu, Daya Guo, Shuo Ren, Junjie Huang, Alexey Svyatkovskiy, Ambrosio Blanco, Colin Clement, Dawn Drain, Daxin Jiang, Duyu Tang, Ge~Li, Lidong Zhou, Linjun Shou, Long Zhou, Michele Tufano, MING GONG, Ming Zhou, Nan Duan, Neel Sundaresan, Shao~Kun Deng, Shengyu Fu, and Shujie LIU. 2021.
\newblock \href {https://openreview.net/forum?id=6lE4dQXaUcb} {Code{XGLUE}: A machine learning benchmark dataset for code understanding and generation}.
\newblock In \emph{Thirty-fifth Conference on Neural Information Processing Systems Datasets and Benchmarks Track (Round 1)}.

\bibitem[{Muennighoff et~al.(2023)Muennighoff, Liu, Zebaze, Zheng, Hui, Zhuo, Singh, Tang, von Werra, and Longpre}]{muennighoff2023octopack}
Niklas Muennighoff, Qian Liu, Armel Zebaze, Qinkai Zheng, Binyuan Hui, Terry~Yue Zhuo, Swayam Singh, Xiangru Tang, Leandro von Werra, and Shayne Longpre. 2023.
\newblock \href {http://arxiv.org/abs/2308.07124} {Octopack: Instruction tuning code large language models}.

\bibitem[{Neelakantan et~al.(2016)Neelakantan, Le, and Sutskever}]{neelakantan2015neural}
Arvind Neelakantan, Quoc~V Le, and Ilya Sutskever. 2016.
\newblock \href {https://arxiv.org/abs/1511.04834} {Neural programmer: Inducing latent programs with gradient descent}.
\newblock In \emph{International Conference on Learning Representations}.

\bibitem[{Ni et~al.(2023)Ni, Yin, Zhao, Riddell, Feng, Shen, Yin, Liu, Yavuz, Xiong, Joty, Zhou, Radev, and Cohan}]{ni2023l2ceval}
Ansong Ni, Pengcheng Yin, Yilun Zhao, Martin Riddell, Troy Feng, Rui Shen, Stephen Yin, Ye~Liu, Semih Yavuz, Caiming Xiong, Shafiq Joty, Yingbo Zhou, Dragomir Radev, and Arman Cohan. 2023.
\newblock \href {http://arxiv.org/abs/2309.17446} {L2ceval: Evaluating language-to-code generation capabilities of large language models}.

\bibitem[{Nijkamp et~al.(2023{\natexlab{a}})Nijkamp, Hayashi, Xiong, Savarese, and Zhou}]{nijkamp2023codegen2}
Erik Nijkamp, Hiroaki Hayashi, Caiming Xiong, Silvio Savarese, and Yingbo Zhou. 2023{\natexlab{a}}.
\newblock \href {http://arxiv.org/abs/2305.02309} {Codegen2: Lessons for training llms on programming and natural languages}.

\bibitem[{Nijkamp et~al.(2023{\natexlab{b}})Nijkamp, Pang, Hayashi, Tu, Wang, Zhou, Savarese, and Xiong}]{nijkamp2023codegen}
Erik Nijkamp, Bo~Pang, Hiroaki Hayashi, Lifu Tu, Huan Wang, Yingbo Zhou, Silvio Savarese, and Caiming Xiong. 2023{\natexlab{b}}.
\newblock \href {https://openreview.net/forum?id=iaYcJKpY2B_} {Codegen: An open large language model for code with multi-turn program synthesis}.
\newblock In \emph{The Eleventh International Conference on Learning Representations}.

\bibitem[{Rahit et~al.(2020)Rahit, Nabil, and Huq}]{10.1007/978-3-030-32520-6_6}
K.~M. Tahsin~Hassan Rahit, Rashidul~Hasan Nabil, and Md~Hasibul Huq. 2020.
\newblock Machine translation from natural language to code using long-short term memory.
\newblock In \emph{Proceedings of the Future Technologies Conference (FTC) 2019}, pages 56--63, Cham. Springer International Publishing.

\bibitem[{Rozière et~al.(2023)Rozière, Gehring, Gloeckle, Sootla, Gat, Tan, Adi, Liu, Remez, Rapin, Kozhevnikov, Evtimov, Bitton, Bhatt, Ferrer, Grattafiori, Xiong, Défossez, Copet, Azhar, Touvron, Martin, Usunier, Scialom, and Synnaeve}]{rozière2023code}
Baptiste Rozière, Jonas Gehring, Fabian Gloeckle, Sten Sootla, Itai Gat, Xiaoqing~Ellen Tan, Yossi Adi, Jingyu Liu, Tal Remez, Jérémy Rapin, Artyom Kozhevnikov, Ivan Evtimov, Joanna Bitton, Manish Bhatt, Cristian~Canton Ferrer, Aaron Grattafiori, Wenhan Xiong, Alexandre Défossez, Jade Copet, Faisal Azhar, Hugo Touvron, Louis Martin, Nicolas Usunier, Thomas Scialom, and Gabriel Synnaeve. 2023.
\newblock \href {http://arxiv.org/abs/2308.12950} {Code llama: Open foundation models for code}.

\bibitem[{Schick et~al.(2023)Schick, Dwivedi-Yu, Dessi, Raileanu, Lomeli, Hambro, Zettlemoyer, Cancedda, and Scialom}]{schick2023toolformer}
Timo Schick, Jane Dwivedi-Yu, Roberto Dessi, Roberta Raileanu, Maria Lomeli, Eric Hambro, Luke Zettlemoyer, Nicola Cancedda, and Thomas Scialom. 2023.
\newblock \href {https://openreview.net/forum?id=Yacmpz84TH} {Toolformer: Language models can teach themselves to use tools}.
\newblock In \emph{Thirty-seventh Conference on Neural Information Processing Systems}.

\bibitem[{Touvron et~al.(2023)Touvron, Martin, Stone, Albert, Almahairi, Babaei, Bashlykov, Batra, Bhargava, Bhosale, Bikel, Blecher, Ferrer, Chen, Cucurull, Esiobu, Fernandes, Fu, Fu, Fuller, Gao, Goswami, Goyal, Hartshorn, Hosseini, Hou, Inan, Kardas, Kerkez, Khabsa, Kloumann, Korenev, Koura, Lachaux, Lavril, Lee, Liskovich, Lu, Mao, Martinet, Mihaylov, Mishra, Molybog, Nie, Poulton, Reizenstein, Rungta, Saladi, Schelten, Silva, Smith, Subramanian, Tan, Tang, Taylor, Williams, Kuan, Xu, Yan, Zarov, Zhang, Fan, Kambadur, Narang, Rodriguez, Stojnic, Edunov, and Scialom}]{touvron2023llama2}
Hugo Touvron, Louis Martin, Kevin Stone, Peter Albert, Amjad Almahairi, Yasmine Babaei, Nikolay Bashlykov, Soumya Batra, Prajjwal Bhargava, Shruti Bhosale, Dan Bikel, Lukas Blecher, Cristian~Canton Ferrer, Moya Chen, Guillem Cucurull, David Esiobu, Jude Fernandes, Jeremy Fu, Wenyin Fu, Brian Fuller, Cynthia Gao, Vedanuj Goswami, Naman Goyal, Anthony Hartshorn, Saghar Hosseini, Rui Hou, Hakan Inan, Marcin Kardas, Viktor Kerkez, Madian Khabsa, Isabel Kloumann, Artem Korenev, Punit~Singh Koura, Marie-Anne Lachaux, Thibaut Lavril, Jenya Lee, Diana Liskovich, Yinghai Lu, Yuning Mao, Xavier Martinet, Todor Mihaylov, Pushkar Mishra, Igor Molybog, Yixin Nie, Andrew Poulton, Jeremy Reizenstein, Rashi Rungta, Kalyan Saladi, Alan Schelten, Ruan Silva, Eric~Michael Smith, Ranjan Subramanian, Xiaoqing~Ellen Tan, Binh Tang, Ross Taylor, Adina Williams, Jian~Xiang Kuan, Puxin Xu, Zheng Yan, Iliyan Zarov, Yuchen Zhang, Angela Fan, Melanie Kambadur, Sharan Narang, Aurelien Rodriguez, Robert Stojnic, Sergey Edunov, and Thomas
  Scialom. 2023.
\newblock \href {http://arxiv.org/abs/2307.09288} {Llama 2: Open foundation and fine-tuned chat models}.

\bibitem[{Wang et~al.(2023{\natexlab{a}})Wang, Li, Qian, Yang, Wang, Shang, Kumar, Tan, Ray, Bhatia, Nallapati, Ramanathan, Roth, and Xiang}]{wang-etal-2023-recode}
Shiqi Wang, Zheng Li, Haifeng Qian, Chenghao Yang, Zijian Wang, Mingyue Shang, Varun Kumar, Samson Tan, Baishakhi Ray, Parminder Bhatia, Ramesh Nallapati, Murali~Krishna Ramanathan, Dan Roth, and Bing Xiang. 2023{\natexlab{a}}.
\newblock \href {https://aclanthology.org/2023.acl-long.773} {{R}e{C}ode: Robustness evaluation of code generation models}.
\newblock In \emph{Proceedings of the 61st Annual Meeting of the Association for Computational Linguistics (Volume 1: Long Papers)}, pages 13818--13843, Toronto, Canada. Association for Computational Linguistics.

\bibitem[{Wang et~al.(2023{\natexlab{b}})Wang, Zhou, Fried, and Neubig}]{wang-etal-2023-execution}
Zhiruo Wang, Shuyan Zhou, Daniel Fried, and Graham Neubig. 2023{\natexlab{b}}.
\newblock \href {https://aclanthology.org/2023.findings-emnlp.89} {Execution-based evaluation for open-domain code generation}.
\newblock In \emph{Findings of the Association for Computational Linguistics: EMNLP 2023}, pages 1271--1290, Singapore. Association for Computational Linguistics.

\bibitem[{Xu et~al.(2022)Xu, Alon, Neubig, and Hellendoorn}]{10.1145/3520312.3534862}
Frank~F. Xu, Uri Alon, Graham Neubig, and Vincent~Josua Hellendoorn. 2022.
\newblock \href {https://doi.org/10.1145/3520312.3534862} {A systematic evaluation of large language models of code}.
\newblock In \emph{Proceedings of the 6th ACM SIGPLAN International Symposium on Machine Programming}, MAPS 2022, page 1–10, New York, NY, USA. Association for Computing Machinery.

\bibitem[{Yin and Neubig(2017)}]{yin-neubig-2017-syntactic}
Pengcheng Yin and Graham Neubig. 2017.
\newblock \href {https://doi.org/10.18653/v1/P17-1041} {A syntactic neural model for general-purpose code generation}.
\newblock In \emph{Proceedings of the 55th Annual Meeting of the Association for Computational Linguistics (Volume 1: Long Papers)}, pages 440--450, Vancouver, Canada. Association for Computational Linguistics.

\bibitem[{Yu et~al.(2024)Yu, Shen, Ran, Zhang, Zhang, Ma, Liang, Li, Wang, and Xie}]{codereval}
H.~Yu, B.~Shen, D.~Ran, J.~Zhang, Q.~Zhang, Y.~Ma, G.~Liang, Y.~Li, Q.~Wang, and T.~Xie. 2024.
\newblock \href {https://doi.ieeecomputersociety.org/} {Codereval: A benchmark of pragmatic code generation with generative pre-trained models}.
\newblock In \emph{2024 IEEE/ACM 46th International Conference on Software Engineering (ICSE)}, pages 417--428, Los Alamitos, CA, USA. IEEE Computer Society.

\bibitem[{Zhang et~al.(2020)Zhang, Kishore, Wu, Weinberger, and Artzi}]{Zhang2020BERTScore:}
Tianyi Zhang, Varsha Kishore, Felix Wu, Kilian~Q. Weinberger, and Yoav Artzi. 2020.
\newblock \href {https://openreview.net/forum?id=SkeHuCVFDr} {Bertscore: Evaluating text generation with bert}.
\newblock In \emph{International Conference on Learning Representations}.

\bibitem[{Zheng et~al.(2023)Zheng, Xia, Zou, Dong, Wang, Xue, Shen, Wang, Wang, Li, Su, Yang, and Tang}]{10.1145/3580305.3599790}
Qinkai Zheng, Xiao Xia, Xu~Zou, Yuxiao Dong, Shan Wang, Yufei Xue, Lei Shen, Zihan Wang, Andi Wang, Yang Li, Teng Su, Zhilin Yang, and Jie Tang. 2023.
\newblock \href {https://doi.org/10.1145/3580305.3599790} {Codegeex: A pre-trained model for code generation with multilingual benchmarking on humaneval-x}.
\newblock In \emph{Proceedings of the 29th ACM SIGKDD Conference on Knowledge Discovery and Data Mining}, KDD '23, page 5673–5684, New York, NY, USA. Association for Computing Machinery.

\bibitem[{Zhou et~al.(2023{\natexlab{a}})Zhou, Sch{\"a}rli, Hou, Wei, Scales, Wang, Schuurmans, Cui, Bousquet, Le, and Chi}]{zhou2023leasttomost}
Denny Zhou, Nathanael Sch{\"a}rli, Le~Hou, Jason Wei, Nathan Scales, Xuezhi Wang, Dale Schuurmans, Claire Cui, Olivier Bousquet, Quoc~V Le, and Ed~H. Chi. 2023{\natexlab{a}}.
\newblock \href {https://openreview.net/forum?id=WZH7099tgfM} {Least-to-most prompting enables complex reasoning in large language models}.
\newblock In \emph{The Eleventh International Conference on Learning Representations}.

\bibitem[{Zhou et~al.(2023{\natexlab{b}})Zhou, Alon, Agarwal, and Neubig}]{zhou-etal-2023-codebertscore}
Shuyan Zhou, Uri Alon, Sumit Agarwal, and Graham Neubig. 2023{\natexlab{b}}.
\newblock \href {https://aclanthology.org/2023.emnlp-main.859} {{C}ode{BERTS}core: Evaluating code generation with pretrained models of code}.
\newblock In \emph{Proceedings of the 2023 Conference on Empirical Methods in Natural Language Processing}, pages 13921--13937, Singapore. Association for Computational Linguistics.

\bibitem[{Zhou et~al.(2023{\natexlab{c}})Zhou, Alon, Xu, Jiang, and Neubig}]{zhou2023docprompting}
Shuyan Zhou, Uri Alon, Frank~F. Xu, Zhengbao Jiang, and Graham Neubig. 2023{\natexlab{c}}.
\newblock \href {https://openreview.net/forum?id=ZTCxT2t2Ru} {Docprompting: Generating code by retrieving the docs}.
\newblock In \emph{The Eleventh International Conference on Learning Representations}.

\end{thebibliography}
\bibliographystyle{acl_natbib}

\appendix

\section{Appendix}
\label{sec:appendix}
\subsection{HumanExtension dataset card}

\begin{table*}[t]
\centering
\begin{tabular}{@{}lc @{}}
\toprule
\multicolumn{1}{c}{Statistics} & \multicolumn{1}{c}{Value} \\ \midrule
Number of examples & 151 \\
Number of testcases per examples & 3.45 \\
Auxiliary function character length per examples (signature only) & 457.05 \\
Auxiliary function character length per examples & 638.97 \\
Target function character length per examples (signature only) & 443.19 \\
Target function character length per examples & 536.44 \\ \bottomrule
\end{tabular}
\caption{Dataset statistics}
\label{tab:my-table}
\end{table*}

\subsubsection{License \& Intended use}
We built our dataset with HumanEval dataset distributed under the MIT license.\footnote{\url{https://huggingface.co/datasets/openai_humaneval}}
The license allows us to modify HumanEval dataset and distribute our new HumanExtension dataset even though the intended use of HumanEval dataset is model evaluation without changing their content.
We plan to release this dataset under the same MIT license.

\subsubsection{Potential risk}
According to the HumanEval dataset card, no personal and sensitive information was contained in the dataset and we confirmed it by inspecting whole problems.
We also conducted a thorough inspection of the newly crafted extended functions and verify that there was no personally identifiable or sensitive information.

\subsubsection{Description}
This dataset contains Python code snippets that contain target functions and the corresponding auxiliary functions that can assist in their implementation.
Additionally, components extracted from abstract syntax parser, e.g., function name and docstring, are included.
The primary use of this dataset is to measure the performance change of generating code depending on the existance of auxiliary functions inside the prompt.
This dataset contains 151 test examples.

\subsubsection{Instruction for crafting problem}
The instruction used for crafting extended functions is as follow.

Dear labelers.

Thanks for participating our job about crafting new Python functions.
Your task is to design a extended function of the given functions by calling the given function or improving their internal mechanism.
There is no constraint about the way for the function extension, but we recommend to read the attached materials \footnote{\url{https://en.wikipedia.org/wiki/Subtyping}}\footnote{\url{https://en.wikipedia.org/wiki/Liskov_substitution_principle}} about the function extension in software engineering fields.
You can pass the examples if you think the function is not appropriate for some reasons, e.g., too specific or too general.

We assumes no responsibility or liability for any potential risk in the labeling process. The information for the creation task is provided on an "as is" basis with no guarantees of completeness, accuracy, usefulness or timeliness.

Sincerely.

\subsection{CodeLLaMAInstruct prompt template}

We follow the template released in their offical github repository \footnote{\url{https://github.com/facebookresearch/codellama}}.
We terminate generation early when eos token or [/PYTHON] is generated.
The following text is the template for generating target function with auxiliary function.

\begin{spverbatim}
<s>[INST] Write a Python function `{auxiliary_function_name}` to solve the following problem:
{auxiliary_function_docstring}
Your code should start with a [PYTHON] tag and end with a [/PYTHON] tag.
[/INST]
[PYTHON]
{auxiliary_function_code}
[/PYTHON]
</s><s>[INST] Write a Python function `{target_function_name}` to solve the following problem:
{target_function_docstring}
Your code should start with a [PYTHON] tag and end with a [/PYTHON] tag.
You can use the above function whenever you needed.
[/INST]
[PYTHON]
{target_function_signature}
\end{spverbatim}

\end{document}